\documentclass[english,12pt]{article}
\usepackage[LGR,T1]{fontenc}
\usepackage[utf8]{inputenc}
\usepackage{geometry}
\usepackage{array}
\usepackage{float}
\usepackage{booktabs}
\usepackage{textcomp}
\usepackage{multirow}
\usepackage{amsmath}
\usepackage{amsthm}
\usepackage{amssymb}
\usepackage{authblk}
\usepackage{sectsty}
\usepackage{graphicx}

\sectionfont{\fontsize{14}{13}\selectfont}
\subsectionfont{\fontsize{12}{12}\selectfont}

\usepackage{blindtext}
\usepackage{caption}
\allowdisplaybreaks

\makeatletter

\DeclareRobustCommand{\greektext}{%
  \fontencoding{LGR}\selectfont\def\encodingdefault{LGR}}
\DeclareRobustCommand{\textgreek}[1]{\leavevmode{\greektext #1}}
\ProvideTextCommand{\~}{LGR}[1]{\char126#1}

\providecommand{\tabularnewline}{\\}

\theoremstyle{plain}
\newtheorem{thm}{\protect\theoremname}

\usepackage{amsmath}
\usepackage[table]{xcolor}
\usepackage{bbm}

\makeatother

\usepackage{babel}
\providecommand{\theoremname}{Theorem}

\begin{document}

\title{Identifying Treatment Effects using Trimmed Means when Data are Missing Not at Random}

\author[1]{Alex Ocampo$^{\dag}$}
\author[2]{Heinz Schmidli}
\author[2]{Peter Quarg}
\author[2]{Francesca Callegari}
\author[1]{Marcello Pagano}

\affil[1]{Department of Biostatistics, Harvard T.H. Chan School of Public Health, Boston, MA}
\affil[2]{Novartis Pharma AG, Basel, Switzerland}
\affil[$\dag$]{Email: ocampo@g.harvard.edu}
\setcounter{Maxaffil}{0}
\renewcommand\Affilfont{\itshape\small}

\date{}
\maketitle
\vspace{-12mm}
\begin{abstract}
Patients often discontinue treatment in a clinical trial because their health condition is not improving. Consequently, the patients still in the study at the end of the trial have better health outcomes on average than the initial patient population would have had if every patient had completed the trial. If we only analyze the patients who complete the trial, then this “missing data problem” biases the estimator of a medication's efficacy because study outcomes are missing not at random (MNAR). One way to overcome this problem - the trimmed means approach for missing data - sets missing values as the worst observed outcome and then trims away a fraction of the distribution from each treatment arm before calculating differences in treatment efficacy (Permutt 2017, Pharmaceutical statistics 16.1: 20-28). In this paper we derive sufficient and necessary conditions for when this approach can identify the average population treatment effect. Numerical studies show the trimmed means approach's ability to effectively estimate treatment efficacy when data are MNAR and missingness is strongly associated with an unfavorable outcome, but trimmed means fail when data are missing at random (MAR). If the reasons for discontinuation in a clinical trial are known analysts can improve estimates with a combination of multiple imputation (MI) and the trimmed means approach when the assumptions of each hold. We compare the methodology to existing approaches using data from a clinical trial for chronic pain. When the assumptions are justifiable, using trimmed means can help identify treatment effects notwithstanding MNAR data.
\end{abstract}

\section{Introduction}

Restricting statistical analysis to patients that complete a clinical
trial can lead to biased results\textsuperscript{1}. Patients with unfavorable
endpoints often discontinue the trial prematurely and thus the remaining
patients no longer provide a truly representative sample of patients,
even if the original sample did. This situation can be viewed as a
missing data problem \textendash{} the endpoints are not measured
for the patients who have left the study and are thus labeled as missing.
There is no consensus on how best to adjust a statistical analysis
for missing data. This is due in part because the reason for missing
data impacts the choice of analytical methods to use. The best understood
situation is when the data are missing at random (MAR)\textsuperscript{2}. This
means that one observes all the data necessary to explain the missingness
in the data - i.e. the missing values themselves did not contribute
to the fact that they are missing. This, of course, is not only an
untestable assumption, but in many settings, it is an unrealistic
one. Its complementary situation, when data are missing not at random
(MNAR), occurs in clinical trials. An example of this is trials of
chronic pain, where subjects are more likely to leave the study if
they experience little or no decrease in pain. Consequently, the outcomes
of those who complete the study differs from those who do not, often
even when accounting for observed information. In these situations,
ignoring the violations of the assumptions underpinning popular missing
data methods designed for MAR data, such as multiple imputation\textsuperscript{3}
and inverse probability weighting\textsuperscript{4}, leads to biases that make
the analyses inadequate and of little scientific value. 

The paucity of data analytic methods for when the data are MNAR may
contribute to the widespread use of inappropriate methods. The National
Research Council (NRC) report on missing data suggests two general
paths forward when data are MNAR: Selection Models and Pattern-Mixture
Models\textsuperscript{5}. Both of these models are limited in that they rely
on parameters that cannot be inferred from the observed data. Therefore,
identification of treatment effects using these models is not possible.
This limitation is emphasized by Little\textsuperscript{6} who highlights that
all MNAR models are subject to a fundamental lack of identification.
The shadow variable approach of Miao \& Tchetgen Tchetgen can identify
causal effects under MNAR\textsuperscript{7}; however, it relies on the presence
of a surrogate outcome - a shadow variable - that is closely related
to the missing outcome and unrelated to why the missing outcome is
unobserved. One may not observe such a variable in the clinical trial
setting. Jump to reference imputation\textsuperscript{8} is a popular approach
to handling MNAR data where the missing outcomes of patients in the
experimental treatment arm are imputed using observed outcomes from
the reference arm. The logic of this approach makes intuitive sense,
in the experimental arm MNAR dropouts have worse outcomes than those
who complete the study. So to reflect this difference one can leverage
the worse outcomes of reference/reference arm for imputations of experimental
dropouts. The main drawback for the MNAR setting is that missing outcomes
in the reference arm are imputed by reference completers which effectively
assumes MAR. If reference arm dropouts are MNAR, then reference imputations
may be optimistic and the treatment effect can be underestimated.
Especially in placebo-controlled trials, one generally expects more
lack of efficacy dropouts in the reference arm which could be problematic
for jump to reference unless initial measurements leveraged for imputations
reveal the poor outcome trajectory.

The \textquotedblleft trimmed means approach\textquotedblright{} for
missing data was first introduced by Li and Permutt\textsuperscript{9}. The method
is designed for settings where missing values can be assumed to be
poor health outcomes that are the continuous endpoints of the study;
precisely the example introduced above. We focus on this method in
the remainder of the paper. The trimmed means approach is simple to
implement in a comparative trial. The final statistic is the arithmetic
mean after trimming the poor performing parts of the distribution
in each arm of the study, after assigning a value to the missing outcomes
that is worse than the worst observed outcome. Applying this approach
only requires the ability to assign a rank to all outcomes and ranks
missing outcomes at the tail end of the distribution prior to trimming.
The trimmed means approach can be extended to include covariates such
as in an ANCOVA and can be applied in mixed models for repeated measures
(MMRM)\textsuperscript{10}. The trimmed-means-approach was designed to estimate
an estimand different than most standard analyses. Instead of estimating
treatment difference in the whole study population, the approach estimates
the treatment difference in a subset of the best performing patients
since those performing poorly are trimmed out of the analysis. As
a result, some of the data is lost, and that is the price one pays
for obtaining an analysis that accommodates MNAR data. The trimmed-means-approach
was evaluated under various missing data generating mechanisms\textsuperscript{11},
which reveals some of its limitations. 

The rest of the paper concentrates on extending the utility of the
trimmed-means-approach for missing data. Section 2 describes the trimmed
means approach and extends its use to a combination with multiple
imputation. Section 3 discusses the estimand and provides a proof
for settings under which the trimmed means approach can identify the
population treatment effect. Section 4 evaluates the finite sample
properties of the approach in a numerical study under various missing
data mechanisms. Section 5 provides recommendations on how to apply
the approach in the context of a randomized clinical trial for a chronic
pain medication. Section 6 concludes the article with a discussion. 

\section{Statistical Methods}

\subsection{The Trimmed Means Estimator}

In 2017, Permutt and Li\textsuperscript{9} introduced the trimmed means estimator.
They provide a thorough motivation of the approach as well as how
to implement it in practice. Mehrotra et al. demonstrate an implementation
of the approach in SAS\textsuperscript{12}. This approach does not rely on any
parametric model or imputation of the missing values; it only depends
on the ability to rank outcomes. The trimmed means approach for missing
data utilizes one sided trimmed means. Consider a continuous outcome
$Y_{i}$ for observation $i$. The one sided trimmed mean is an L-estimator
defined as: 

\[
\hat{\mu}_{T}=\frac{1}{n_{T}}\sum_{i:Y_{i}>\hat{F}^{-1}(\alpha)}^{n}Y_{i}
\]

The above is simply the average of the observations that fall above
the quantile $\hat{F}^{-1}(\alpha)$ of the empirical distribution
function of $Y_{i}$ where $\alpha$ represents the proportion of
trimmed outcomes. For example, $\alpha=0.3$ represents trimming away
the bottom 30\% of the distribution. Here, $n_{T}=\sum_{i=1}^{n}\mathbbm{1}(Y_{i}>\hat{F}^{-1}(\alpha))$
is the sample size after trimming. The expectation of the trimmed
mean is the population trimmed mean:

\[
\mu_{T}=E[Y|Y>F^{-1}(\alpha)]=\frac{1}{1-\alpha}\int_{F^{-1}(\alpha)}^{\infty}yf(y)dy
\]

Note that for this example above, poor outcomes correspond to low
values of $Y$, so the lower part of the distribution is trimmed.
This could be switched based on the clinical context. For example,
in pain trials a decrease in pain intensity is a good outcome, so
one would rather trim away the upper part of the distribution. 

To implement this approach, first consider three observed variables:
$A$ a binary indicator for treatment, $R$ a binary missing data
indicator, and the clinical outcome $Y$ which is a continuous value
when $R=1$ and is missing when $R=0$. Operationally, the first step
of the trimmed means approach is to remove the missing data by ranking
all missing outcomes as slightly worse than the poorest observed outcome
in the trial. To do so one defines a new outcome denoted by $U$ for
each subject $i$:

\[
U_{i}=\begin{cases}
Y_{i} & \text{if}\;R_{i}=1\\
min(\mathbf{Y})-\epsilon & \text{if}\;R_{i}=0
\end{cases}\text{for}\;\epsilon>0
\]

Note the above corresponds to low values representing poor outcomes.
If high values of the outcome reflected poor values then if $R=0$
the missing outcomes would be set to $max(\mathbf{Y})+\epsilon$.
After ranking, $\alpha$ proportion of each treatment arm is trimmed
away from the end of the distribution of $U$ associated with poor
outcomes. The analyst has some flexibility in determining $\alpha$,
the proportion of data trimmed from each distribution, before calculating
the treatment effect. This value can be fixed a priori with a value
of $\alpha$ chosen that anticipates the amount of missing data. Alternatively,
trimming can be adaptive and chosen to be the maximum between the
proportions of missing data in each arm. Thus the values that the
missing observations were set to are never actually used in the analysis,
but do serve the important function of informing the quantiles at
which each distribution is to be trimmed. After trimming, then calculate
the mean of the remaining $100\times(1-\alpha)\%$ observations in
each arm $\hat{\mu}_{T1}$ and $\hat{\mu}_{T0}$. The final estimate
is obtained by taking the difference of these trimmed means between
each arm: $\hat{\mu}_{T\Delta}=\hat{\mu}_{T1}-\hat{\mu}_{T0}$.

Inference can be conducted via a permutation test that conditions
on the observed data and randomly permutes the treatment assignments.
The resulting permutation distribution of treatment differences formed
under the null hypothesis can be used to determine significant differences
and confidence intervals. To reject the null hypothesis of no treatment
difference, the point estimate should fall above the upper 2.5th percentile
of the permutation distribution. If $\hat{\mu}_{T\Delta}$ is the
treatment difference calculated after trimming, a 95\% confidence
interval can be constructed by adding the 97.5th and 2.5th percentiles
of the permutation distribution to $\mu_{T\Delta}$. This can be generalized
for any significance level $\gamma$ such that $\left(\hat{\mu}_{T\Delta}-y_{\gamma/2},\hat{\mu}_{T\Delta}+y_{1-\gamma/2}\right)$
yields a $(1-\gamma)\%$ confidence interval where $y_{\gamma}$ is
the $\gamma$ percentile of the permutation distribution. Since these
confidence intervals are constructed using the permutation distribution
generated under the null hypothesis, the intervals will be conservative
when the null hypothesis is false\textsuperscript{13}.

\subsection{Combining Trimmed Means with Multiple Imputation}

In well conducted clinical trials, the reason for study discontinuation
is collected for each patient who drops out of the study. Treating
certain types of dropout as poor outcomes and ranking them at the
low end of the distribution, as the trimmed means approach does, would
lead to biases. Knowing the reason for dropping out of a study, and
using that information, should lead to more precise analyses. To that
end, consider the expanded indicator:

\[
R=\begin{cases}
r_{1} & \text{if}\;Y\;\text{observed}\\
r_{2} & \text{if MAR or MCAR}\\
r_{3} & \text{if MNAR}
\end{cases}
\]

Assume the complete data of $Y$ is partitioned into the observed
and missing components as follows $Y=\left[Y_{r_{1}},Y_{r_{2}},Y_{r_{3}}\right]$
which denote the observed, missing at random, and missing not at random
components of $Y$ respectively. Here we propose imputing $Y_{r_{2}}$
and trimming $Y_{r_{3}}$. We can perform multiple imputation of $Y_{r_{2}}$
when the conditional distribution $f(Y_{r_{2}}|Y_{r_{1}}=y_{r_{1}},A,\mathbf{X})$
is a valid imputation model given the MAR assumption. Here $y_{r_{1}}$
denotes the observed outcomes, $A$ is the treatment assignment, and
$\mathbf{X}$ is a matrix of auxiliary covariates that may or may
not be available and of use for imputations. Using this conditional
distribution, one can draw $m$ samples for the MAR and MCAR data
$Y_{r_{2}}^{(1)},Y_{r_{2}}^{(2)},\dots,Y_{r_{2}}^{(m)}$ to derive
a set of data that is now complete for $Y_{r_{2}}$ where missing
values $Y_{r_{3}}$ remain. Let $\hat{\mu}_{T\Delta}=\hat{\mu}_{T\Delta}\left(Y_{r_{1}},Y_{r_{2}},\mathbbm{1}(R=r_{3})\right)$
denote the trimmed means statistic given that complete data on $Y_{r_{2}}$
were available. Note we do not need to observe $Y_{r_{3}}$ since
the trimmed means approach will trim these observations out of the
analysis. Multiple imputation relies on the asymptotically normal
distribution of $\hat{\mu}_{T\Delta}$, which applies to the trimmed
mean\textsuperscript{14}. Since data on $Y_{r_{2}}$ are missing, the imputed
data are utilized to calculate trimmed means estimates of the form
$\hat{\mu}_{T\Delta}^{(\ell)}=\hat{\mu}_{T\Delta}\left(Y_{r_{1}},Y_{r_{2}}^{(\ell)},\mathbbm{1}(R=r_{3})\right)$
for the $m$ imputed datasets. Lastly, we Rubin's rules {[}3{]} to
summarize the results of the trimmed means applied to each partially
imputed dataset.
\begin{align*}
\bar{\mu}_{T\Delta} & =\frac{1}{m}\sum_{\ell=1}^{m}\mu_{T\Delta}^{(\ell)}=\frac{1}{m}\sum_{\ell=1}^{m}\mu_{T1}^{(\ell)}-\mu_{T0}^{(\ell)}\\
Var(\bar{\mu}_{T\Delta}) & =\frac{1}{m}\sum_{\ell=1}^{m}(\sigma^{(\ell)})^{2}+\left(1+\frac{1}{m}\right)\left(\frac{1}{m-1}\sum_{\ell=1}^{m}\left(\mu_{\Delta}^{(\ell)}-\bar{\mu}_{T\Delta}\right)^{2}\right)
\end{align*}

Where $\sigma^{(\ell)}$ is the estimated standard error of the trimmed
means estimate in the $\ell$th imputed dataset. This combination
approach is only valid if all unobserved values of $Y_{r_{3}}$ fall
below the quantile of the distributions that are trimmed, a condition
discussed in the subsequent section.

\section{Properties of the Estimand }

\subsection{Equivalence to the Population Treatment Effect}

We focus on a randomized clinical trial, where the estimand of interest
is the treatment difference in the population means of a clinical
endpoint. Consider counterfactual outcomes $Y_{a}$ where $a\in{0,1}$
indicates potential treatment assignments. In addition, $A\in{0,1}$
is the binary indicator for observed treatment in the trial. Denote
$U_{a}$ as the counterfactual version of the composite outcome defined
in section 2. The trimmed means estimand is most similar to a composite
estimand, using the terminology of the ICH E9 addendum \textsuperscript{15}. The
trimmed means estimand in counterfactual notation is $E[U_{1}|U_{1}>F_{U_{1}}^{-1}(\alpha)]-E[U_{0}|U_{0}>F_{U_{0}}^{-1}(\alpha)]$
where $F_{U_{a}}^{-1}(\alpha)$ represents the inverse cdf of the
counterfactual distributions of $U_{a}$ evaluated at $\alpha$. This
is a unique estimand defined for a sub-population of the trial, interpreted
as the difference between treatments in endpoint means in the best
$100\times(1-\alpha)\%$ of patients from each arm. This is not, however,
the treatment effect among all randomized patients.

The advantage of the trimmed means approach\textquoteright s composite
estimand is that it gives us a strategy for handling the missing data;
however, the main drawback of using a composite estimand is that the
clinical relevance can be unclear. In may be difficult and unfamiliar
to interpret an estimated treatment effect in the best $100\times(1-\alpha)\%$
of patients as compared to an estimate of efficacy for all patients
in a particular indication. The estimand for treatment efficacy in
the population from which all randomized patients are drawn is the
difference in counterfactual means $E[Y_{1}]-E[Y_{0}]$, however using
the trimmed means approach only the difference in trimmed means of
the composite outcomes $U_{1}$ and $U_{0}$ are estimated. Herein,
the sufficient and necessary conditions under which the trimmed means
estimand and the population estimand for treatment effect are identical
are formalized.
\begin{thm}
If the outcomes among the treated and untreated are identically distributed
relative to a shift and all unobserved values fall below the trim,
then the treatment difference estimated by the trimmed means approach
is equivalent to the treatment difference in the population, i.e.:
\[
E[U_{1}|U_{1}>F_{U_{1}}^{-1}(\alpha)]-E[U_{0}|U_{0}>F_{U_{0}}^{-1}(\alpha)]=E[Y_{1}]-E[Y_{0}]
\]
\end{thm}

The proof is given in the appendix. 

$\;$

Theorem 1 uses the following two conditions in order to prove the
equality of the trimmed means estimand and treatment difference in
the whole population.
\begin{enumerate}
\item \textbf{Location family assumption.} The distribution of potential
outcomes had the patient taken the experimental treatment $Y_{1}\sim f_{1}(y)$
is in the same location family as the distribution of potential outcomes
had the patient taken the reference treatment $Y_{0}\sim f_{0}(y)$.
Consider some constant $\Delta$ then:
\[
f_{0}(y)=f_{1}(y+\Delta)
\]
\item \textbf{Strong MNAR assumption. }All missing values fall below the
point at which the distributions are trimmed. Explicitly, the strong
MNAR assumption states: 
\[
Y_{a}|R_{a}=0<F_{a}^{-1}(\alpha)
\]
\end{enumerate}
The strong MNAR assumption ensures that the composite outcome $U_{a}$
is trimmed at the same value as $Y_{a}$ for all percentiles above
the maximum rate of missing data between the two arms, i.e. $F_{U_{a}}^{-1}(\alpha)=F_{a}^{-1}(\alpha)\forall\alpha:\alpha>Pr[R_{a}=0]$.
It also guarantees that the untrimmed distribution of $U_{a}$ is
identical to that of $Y_{a}$.

If $Y_{1}$ and $Y_{0}$ can be identified from the observed data,
then the trimmed means approach can estimate the causal estimand of
treatment effect in the population given the above two assumptions
as shown:
\begin{align*}
E[Y_{1}]-E[Y_{0}] & =E[Y_{1}|Y_{1}>F_{1}^{-1}(\alpha)]-E[Y_{0}|Y_{0}>F_{0}^{-1}(\alpha)]\;\;\;\;(1)\\
 & =E[U_{1}|U_{1}>F_{U_{1}}^{-1}(\alpha)]-E[U_{0}|U_{0}>F_{U_{0}}^{-1}(\alpha)]\;\;\;(2)\\
 & =E[U_{1}|U_{1}>F_{U_{1}}^{-1}(\alpha),A=1]-E[U_{0}|U_{0}>F_{U_{0}}^{-1}(\alpha),,A=0]\\
 & =E[U|U>F_{U_{1}}^{-1}(\alpha),A=1]-E[U_{0}|U_{0}>F_{U_{0}}^{-1}(\alpha),A=0]
\end{align*}

The location family assumption makes the average difference of the
entire population of counterfactuals equivalent to the difference
in the sub-population of counterfactuals that are not trimmed (1).
The strong MNAR assumption makes the trimmed means of the counterfactuals
equivalent to the trimmed means of the composite outcome (2). Then
randomization and consistency allow us to identify the counterfactuals
from the observed data\textsuperscript{16}. Note that using Theorem 1 only the
difference in the counterfactual means can be recovered, albeit with
a smaller sample size than if there were no MNAR data. One cannot
accurately estimate the marginal means of $Y_{1}$ and $Y_{0}$ in
the presence of MNAR data.
\begin{thm}
If the difference in untrimmed means between two counterfactual distributions
is equivalent to the difference in one sided trimmed means for all
percentiles,
\[
E[Y_{1}|Y_{1}>F_{1}^{-1}(\alpha)]-E[Y_{0}|Y_{0}>F_{0}^{-1}(\alpha)]=E[Y_{1}]-E[Y_{0}]=\Delta\;\forall\alpha(0,1)
\]
then the counterfactual distributions are a location shift of one
another
\[
f_{0}(y)=f_{1}(y+\Delta)
\]
\end{thm}

Proof is given in the appendix.

$\;$

Theorem 2 demonstrates that using the trimmed means approach to estimate
the population treatment effect is only relevant for treatments with
an additive effect. For all possible $\alpha$, the difference in
trimmed means and the population mean are equivalent if and only if
the distributions being compared are a location shift of one another.
There are conditions where the difference in trimmed means and population
means are equivalent when the strong MNAR assumption is not true (i.e.
the MCAR case). Thus, theorem 2 reveals that the location family assumption
is a sufficient and necessary condition, while the strong MNAR assumption
is only a sufficient condition for the equivalence of the estimands.

\subsection{Intercurrent Events}

Theorem 1 extends the utility of the trimmed means approach by proving
under what assumptions one can estimate the estimand representing
the treatment effect based on all randomized patients rather than
a subset of the best performing patients. It is important to discuss
how this result fits into the estimand framework outlined in the ICH
E9 addendum\textsuperscript{15}. The trimmed means approach is a statistical
analysis that specifies how to deal with missing data, not necessarily
a particular strategy to deal with intercurrent events (IE), which
may often, but do not deterministically, lead to missing data. The
IE strategy depends on how one handles data after observing an IE
and is one of the four components in defining the estimand. 

As discussed above, the trimmed means can be used as a composite approach,
whereby IE that lead to missing data - and potentially others - are
ranked poorly and trimmed out of the analysis. The resulting composite
estimand can be thought of as a measure of treatment difference that
is penalized by the amount of dropout in each arm. This penalty comes
by placing the IE towards the poor end of the distribution irrespective
of if the unobserved data are MAR or MNAR. In other words, they are
ranked as the worst outcome even if their outcome would not have been
poor had they continued in the trial. This type of estimand seems
ideal for IE such as adverse events that outweigh the benefit of treatment
or death. To estimate this composite estimand, the location family
and strong MNAR assumptions do not need to hold. Should these assumptions
be realistic however, the opportunity arises to use the trimmed means
approach to estimate two other types of estimands: 1) the hypothetical
and 2) intention to treat (ITT) estimands.

The hypothetical estimand postulates what would have happened had
the intercurrent event not occurred and the patient remained on treatment
for the duration of the trial. Even if post IE data is collected,
it is discarded and treated as missing data. For the ITT estimand
the IE is irrelevant, and one is interested in data that occur after
the patient discontinues treatment. If this post IE data is not missing,
it is used as a valid endpoint in the analysis. Both the ITT and hypothetical
estimands are estimable using the trimmed means approach if the strong
MNAR and location family assumptions of Theorem 1 hold. The difference
between using the trimmed means approach for the hypothetical and
ITT estimands is that the assumptions are made on different counterfactuals
that are determined by whether or not treatment is continued post-IE.
It seems more likely that the location family assumption in particular
would hold for the hypothetical estimand, especially when the drug
has an additive effect. The strong MNAR assumption seems more likely
for the ITT estimand, but is ultimately based on the process generating
the missing data. In addition, if information is collected on the
reasons for missing data, and if some of these missing data can be
assumed to be MAR and other strong MNAR then one could use a combination
of multiple imputation and trimmed means to estimate these estimands. 

\section{Numerical Studies}

\subsection{Simulation Objectives}

Numerical studies herein evaluate the finite sample properties of
the trimmed means approach in estimating treatment efficacy under
various missing data generating mechanisms. The simulation presented
is motivated by the design of Wang et al. \textsuperscript{11}. This earlier work
is extended in a number of ways. Firstly, a comparison to MI under
the various missing data generating mechanisms of the simulation is
demonstrated. Additionally, when there exists a mixture of missing
data types the combination approach is evaluated and compared to applying
the trimmed means approach and MI globally. Furthermore, the relationship
between bias and violation of the strong MNAR assumption of theorem
1 is considered under different MNAR scenarios. Lastly, the choice
of $\alpha$ is explored. The comparison of trimmed means to MI as
well as the combination approach would not be possible without Theorem
1 because it demonstrates that the approaches can estimate the same
estimand. Data are imputed using the mice package in R, which leverages
the same methodology used by PROC MI in SAS\textsuperscript{17}.

\subsection{Simulation Design}

We design the study using four different ways to generate the missing
data: a) Missing Completely at Random (MCAR), b) Missing at Random
(MAR), c) Missing Not at Random (MNAR), and d) a mixture of all three
types. Figure 1 displays the causal diagrams for these simulation
designs using the $m$-graphs of Pearl\textsuperscript{18}.

\begin{figure}[H]
    \centering
    \vspace*{0mm}\hspace*{0cm}\includegraphics[scale = 0.35]{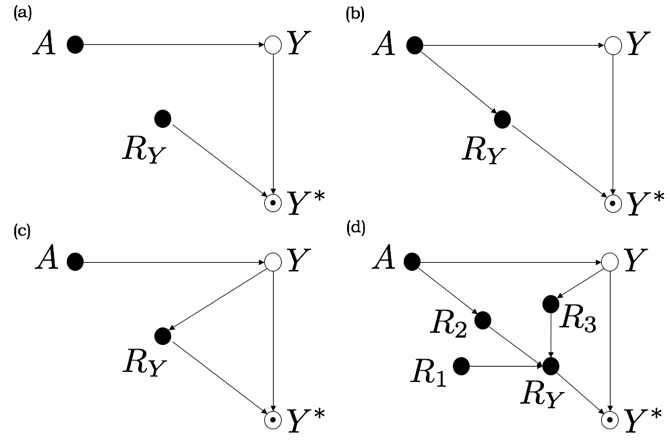}
    \caption{$m$-graphs for scenarios A, B, C, and D of the numerical study}
    \label{fig:my_label}
\end{figure}

 In these diagrams $A$ represents a binary indicator for treatment,
$Y$ a continuous clinical endpoint of the study, and $R_{Y}$ a binary
indicator variable that is equal to 1 if $Y$ is observed and 0 if
$Y$ is missing. Variable $Y^{*}$ is the observed outcome, and its
partially filled in node on the graph indicates that it has some missing
values. Filled in nodes represent fully observed variables (i.e. $A$
and $R_{Y}$). Nodes that are not filled in represent unobserved variables
(i.e. $Y$). The arrows in these graphs make explicit the assumptions
about which variables have a causal effect on missingness.

We use a study sample size of $N=100$ ($n=50$ per treatment arm)
in each of the four scenarios. Each scenario was replicated $K=5000$
times. The $\alpha$ parameter that determines which percentile to
trim in the analysis is chosen adaptively, unless stated otherwise.
The upper part of the distribution is trimmed, corresponding to lower
values reflecting better outcomes. The underlying model for the continuous
outcome remains the same in all simulations:
\vspace{-1mm}
\[
Y=\beta_{0}+\beta_{A}A+\epsilon
\]

$Y$ is the continuous outcome variable and $A$ is the binary variable
representing experimental treatment if 1 and reference treatment if
0. Here, the error term is normally distributed $\epsilon\sim N(0,\sigma^{2})$.
The goal is to estimate $\beta_{A}$ , the difference of the means
between treatments. In all scenarios, the values of the parameters
for the outcome model are $\beta_{0}=-1,\beta_{A}=-1,\sigma=1.5$.
We chose $\sigma=1.5$ to obtain a benchmark \textasciitilde{}90\%
power in a one-sided t-test when there is no missing data. 

The missing data in outcome $Y$ were generated via the following
logit model:
\vspace{-1mm}
\[
Pr(R_{Y}=1|A,Y)=logit^{-1}(\alpha_{0}+\alpha_{A}A+\alpha_{Y}Y)
\]

Where $R_{Y}$ is the binary variable indicating that $Y$ has been
observed if equal to 1. In this model, setting parameters $\alpha_{A}=\alpha_{Y}=0$
corresponds to MCAR because the missing values are unrelated to treatment
or outcome. Setting parameter $\alpha_{Y}=0$ corresponds to MAR because
the missing values are only dependent on the observed values and not
the unobserved outcome. If $\alpha_{Y}\neq0$ then the model represents
an MNAR missing data generating mechanism. 

\subsection{Simulation Results}

\subsection*{(a) MCAR}

In the MCAR setting the $\alpha_{0}$ parameter is set to values of
2.94, 2.20, 1.74, and 1.39 to induce missing data rates of 5, 10,
15, and 20 percent while keeping $\alpha_{A}=\alpha_{Y}=0$. The missing
data rates are the same on average in each arm since unobserved outcomes
are completely random.

\begin{table}[H]
\begin{centering}
\begin{tabular}{ccccccc}
\toprule 
\multicolumn{2}{c}{Missing Rate, \%} & \multicolumn{5}{c}{Trimmed Means}\tabularnewline
\midrule 
$A=1$ & $A=0$ & Exp & Ref & Diff (\% bias) & Coverage & Power\tabularnewline
\midrule
\midrule 
5 & 5 & -2.17 & -1.17 & -1.00 (0\%) & 0.96 & 0.86\tabularnewline
\midrule 
10 & 10 & -2.16 & -1.16 & -1.00 (0\%) & 0.96 & 0.83\tabularnewline
\midrule 
15 & 15 & -2.15 & -1.16 & -1.00 (0\%) & 0.96 & 0.80\tabularnewline
\midrule 
20 & 20 & -2.39 & -1.39 & -1.00 (0\%) & 0.96 & 0.70\tabularnewline
\bottomrule
\end{tabular}
\par\end{centering}
\caption{Trimmed Means with MCAR}
\end{table}

Under a completely random missing data generating mechanism (MCAR),
the trimmed means approach estimates the true treatment difference
without bias and with appropriate coverage even as the proportion
of data missing varies (Table 1). As expected, power decreases as
the amount of data not trimmed decreases. MI performed similarly in
that bias and coverage were accurate (Table 2). However, as the amount
of missing data increases, power does not deteriorate as quickly using
MI than when using trimmed means. This is because the trimmed means
approach performs inference on the subset of the observations post-trimming
and thus uses a smaller effective sample size.

\begin{table}[H]
\begin{centering}
\begin{tabular}{ccccccc}
\toprule 
\multicolumn{2}{c}{Missing Rate, \%} & \multicolumn{5}{c}{Multiple Imputation (MI)}\tabularnewline
\midrule 
$A=1$ & $A=0$ & Exp & Ref & Diff (\% bias) & Coverage & Power\tabularnewline
\midrule
\midrule 
5 & 5 & -2.00 & -1.00 & -1.00 (0\%) & 0.94 & 0.89\tabularnewline
\midrule 
10 & 10 & -2.00 & -1.00 & -1.00 (0\%) & 0.94 & 0.87\tabularnewline
\midrule 
15 & 15 & -1.99 & -1.00 & -0.99 (1\%) & 0.94 & 0.84\tabularnewline
\midrule 
20 & 20 & -1.99 & -1.01 & -0.99 (1\%) & 0.94 & 0.82\tabularnewline
\bottomrule
\end{tabular}
\par\end{centering}
\caption{Multiple Imputation with MCAR}
\end{table}

\subsection*{(b) MAR}

In the MAR setting, we first set the $\alpha_{A}$ parameter to values
of -8.61, -8.27, -7.80, and -7.06 to induce missing data rates of
20, 15, 10, and 5 percent in the experimental arm while keeping $\alpha_{Y}=0$
and $\alpha_{0}=10$ in order to maintain all outcomes observed in
the reference arm. Next, we set $\alpha_{Y}=0$ and $\alpha_{A}=10$
in order to fully observe outcomes in the experimental arm while varying
$\alpha_{0}$ to values of 2.94, 2.20, 1.73, and 1.39 to induce missing
data rates of 5, 10, 15, and 20 percent in the reference arm.

\begin{table}[H]
\begin{centering}
\begin{tabular}{ccccccc}
\toprule 
\multicolumn{2}{c}{Missing Rate, \%} & \multicolumn{5}{c}{Trimmed Means}\tabularnewline
\midrule 
$A=1$ & $A=0$ & Exp & Ref & Diff (\% bias) & Coverage & Power\tabularnewline
\midrule
\midrule 
20 & 0 & -1.99 & -1.51 & -0.48 (52\%) & 0.74 & 0.22\tabularnewline
\midrule 
15 & 0 & -2.00 & -1.41 & -0.59 (40\%) & 0.81 & 0.37\tabularnewline
\midrule 
10 & 0 & -2.00 & -1.28 & -0.72 (28\%) & 0.88 & 0.54\tabularnewline
\midrule 
5 & 0 & -2.00 & -1.15 & -0.84 (16\%) & 0.93 & 0.74\tabularnewline
\midrule 
0 & 5 & -2.16 & -1.01 & -1.15 (-15\%) & 0.95 & 0.95\tabularnewline
\midrule 
0 & 10 & -2.28 & -1.00 & -1.28 (-28\%) & 0.90 & 0.97\tabularnewline
\midrule 
0 & 15 & -2.40 & -1.00 & -1.40 (-20\%) & 0.85 & 0.98\tabularnewline
\midrule 
0 & 20 & -2.51 & -1.00 & -1.51 (-51\%) & 0.77 & 0.99\tabularnewline
\bottomrule
\end{tabular}
\par\end{centering}
\caption{Trimmed Means with MAR}
\end{table}

As expected, the trimmed means estimator is biased in all scenarios
when the missing data is truly MAR (Table 3). The bias increases when
the fraction of missing data increases. The direction of the bias
is positive when the reference arm has more missing data and negative
when the active arm has more missing data. This directionality of
the bias has an impact on power, with more MAR data in the active
arm leading to a drastic decrease in power and more MAR data in the
reference arm causing unreasonably high power. The trimming is directional
as all missing values are placed at the poor end of each respective
treatment distribution when in reality under MAR they come from all
areas of the distribution. MI obtains valid estimation in this setting
as it was designed explicitly for situations where data are MAR (Table
4). 

\begin{table}[H]
\begin{centering}
\begin{tabular}{ccccccc}
\toprule 
\multicolumn{2}{c}{Missing Rate, \%} & \multicolumn{5}{c}{Multiple Imputation (MI)}\tabularnewline
\midrule 
$A=1$ & $A=0$ & Exp & Ref & Diff (\% bias) & Coverage & Power\tabularnewline
\midrule
\midrule 
20 & 0 & -1.99 & -1.00 & -0.99 (-1\%) & 0.94 & 0.86\tabularnewline
\midrule 
15 & 0 & -2.00 & -1.00 & -1.00 (0\%) & 0.94 & 0.88\tabularnewline
\midrule 
10 & 0 & -2.00 & -1.00 & -1.00 (0\%) & 0.95 & 0.89\tabularnewline
\midrule 
5 & 0 & -1.99 & -1.00 & -1.00 (0\%) & 0.95 & 0.91\tabularnewline
\midrule 
0 & 5 & -2.00 & -1.00 & -1.00 (0\%) & 0.94 & 0.90\tabularnewline
\midrule 
0 & 10 & -2.00 & -1.00 & -1.00 (0\%) & 0.95 & 0.90\tabularnewline
\midrule 
0 & 15 & -2.00 & -1.01 & -1.00 (0\%) & 0.95 & 0.88\tabularnewline
\midrule 
0 & 20 & -2.00 & -1.01 & -1.00 (0\%) & 0.94 & 0.86\tabularnewline
\bottomrule
\end{tabular}
\par\end{centering}
\caption{Multiple Imputation with MAR}
\end{table}

\subsection*{(c) MNAR}

In the MNAR setting, the $\alpha_{Y}$ parameter was set to values
of -1, -2.5, -5, and -10 causing higher values of $Y$ to be more
likely to be missing while keeping $\alpha_{0}=2.85$ and $\alpha_{A}=0$.
Here, $\alpha_{Y}$ is negative because a decrease in $Y$ reflects
a better outcome. This setup induces missing data rates in the experimental
vs reference arms of 2 vs 5, 3 vs 10, 5 vs 15, and 7 vs 20 respectively.
The missing data are not simulated strictly as strong MNAR but a general
MNAR missing data mechanism.

\begin{table}[H]
\begin{centering}
\begin{tabular}{ccccccc}
\toprule 
\multicolumn{2}{c}{Missing Rate, \%} & \multicolumn{5}{c}{Trimmed Means}\tabularnewline
\midrule 
$A=1$ & $A=0$ & Exp & Ref & Diff (\% bias) & Coverage & Power\tabularnewline
\midrule
\midrule 
2 & 5 & -2.14 & -1.11 & -1.04 (-4\%) & 0.96 & 0.90\tabularnewline
\midrule 
3 & 10 & -2.28 & -1.26 & -1.02 (-2\%) & 0.96 & 0.90\tabularnewline
\midrule 
5 & 15 & -2.41 & -1.41 & -1.00 (0\%) & 0.96 & 0.90\tabularnewline
\midrule 
7 & 20 & -2.51 & -1.51 & -1.00 (0\%) & 0.95 & 0.89\tabularnewline
\bottomrule
\end{tabular}
\par\end{centering}
\caption{Trimmed Means with MNAR}
\end{table}

The trimmed means approach is fairly unbiased, obtains ideal coverage,
and maintains its power in the MNAR setup as the amount of missing
data increases (Table 5). While the marginal means in each arm are
biased, the means in each arm increase at equal rates and keep the
estimate of their difference unbiased. Multiple imputation increases
bias, reduces coverage of the true effect, and loses power as the
fraction of MNAR data increases (Table 6).

\begin{table}[H]
\begin{centering}
\begin{tabular}{ccccccc}
\toprule 
\multicolumn{2}{c}{Missing Rate, \%} & \multicolumn{5}{c}{Multiple Imputation (MI)}\tabularnewline
\midrule 
$A=1$ & $A=0$ & Exp & Ref & Diff (\% bias) & Coverage & Power\tabularnewline
\midrule
\midrule 
2 & 5 & -2.10 & -1.10 & -1.00 (0\%) & 0.95 & 0.92\tabularnewline
\midrule 
3 & 10 & -2.14 & -1.26 & -0.89 (11\%) & 0.93 & 0.88\tabularnewline
\midrule 
5 & 15 & -2.19 & -1.41 & -0.78 (22\%) & 0.87 & 0.81\tabularnewline
\midrule 
7 & 20 & -2.22 & -1.52 & -0.70 (30\%) & 0.79 & 0.74\tabularnewline
\bottomrule
\end{tabular}
\par\end{centering}
\caption{Multiple Imputation with MNAR}
\end{table}

\subsection*{(d) Mixture: MCAR, MAR, and MNAR}

Having a mixture of reasons for missing data reflects the information
one would have in a closely monitored clinical trial. In many trials,
data are missing for a combination of reasons such as lack of efficacy,
intolerability, and administrative reasons. In order to generate such
data the deletion strategies used in the previous three sections are
combined. MNAR data (R3) were deleted first at rates of 2 vs 5, 3
vs 10, 5 vs 15, and 7 vs 20 in the experimental vs reference arms
respectively. MAR data (R2) were then generated in the experimental
group at rates of 23, 17, 10, and 3. MCAR data (R1) were generated
at a rate of 5 percent in each arm. Overall, the missing data rates
in the four mixture scenarios in the experimental vs reference arms
are 10 vs 30, 15 vs 25, 20 vs 20, and 15 vs 25 respectively.

In the mixture setting, the performance of trimmed means applied globally
is directly related to the proportion of MAR missing data (Table 7).
This fraction of MAR data decreases across the four scenarios and
consequently bias, coverage, and power improve across the scenarios.
Contrarily, MI applied globally performs well with a large fraction
of MAR data and its performance weakens as the proportion of MNAR
data increases (Table 8). The combination of trimmed means and MI
exhibits improved bias, coverage, and power as compared to each method
applied individually (Table 9). Bias is at most 3\%, no matter the
variation in the fraction of missing data due to MAR and MNAR. Coverage
and Power are near the optimal 0.95 and 0.90. 

\begin{table}[H]
\begin{centering}
\begin{tabular}{cccccccccc}
\toprule 
 & \multicolumn{4}{c}{Missing Rate, \%} & \multicolumn{5}{c}{Trimmed Means}\tabularnewline
\midrule 
Trt & R1 & R2 & R3 & Overall & Exp & Ref & Diff (\% bias) & Coverage & Power\tabularnewline
\midrule
\midrule 
$A=1$ & 5 & 23 & 2 & 30 & \multirow{2}{*}{-2.05} & \multirow{2}{*}{-1.65} & \multirow{2}{*}{-0.40 (60\%)} & \multirow{2}{*}{0.71} & \multirow{2}{*}{0.13}\tabularnewline
\cmidrule{1-5} 
$A=0$ & 5 & 0 & 5 & 10 &  &  &  &  & \tabularnewline
\midrule 
$A=1$ & 5 & 17 & 3 & 25 & \multirow{2}{*}{-2.12} & \multirow{2}{*}{-1.55} & \multirow{2}{*}{-0.56 (44\%)} & \multirow{2}{*}{0.80} & \multirow{2}{*}{0.30}\tabularnewline
\cmidrule{1-5} 
$A=0$ & 5 & 0 & 10 & 15 &  &  &  &  & \tabularnewline
\midrule 
$A=1$ & 5 & 10 & 5 & 20 & \multirow{2}{*}{-2.24} & \multirow{2}{*}{-1.47} & \multirow{2}{*}{-0.77 (23\%)} & \multirow{2}{*}{0.90} & \multirow{2}{*}{0.56}\tabularnewline
\cmidrule{1-5} 
$A=0$ & 5 & 0 & 15 & 20 &  &  &  &  & \tabularnewline
\midrule 
$A=1$ & 5 & 3 & 7 & 15 & \multirow{2}{*}{-2.48} & \multirow{2}{*}{-1.54} & \multirow{2}{*}{-0.93 (7\%)} & \multirow{2}{*}{0.94} & \multirow{2}{*}{0.78}\tabularnewline
\cmidrule{1-5} 
$A=0$ & 5 & 0 & 20 & 25 &  &  &  &  & \tabularnewline
\bottomrule
\end{tabular}
\par\end{centering}
\caption{Trimmed Means with a Mixture of Missing Data Types}
\end{table}

\begin{table}[H]
\begin{centering}
\begin{tabular}{cccccccccc}
\toprule 
 & \multicolumn{4}{c}{Missing Rate, \%} & \multicolumn{5}{c}{Multiple Imputation (MI)}\tabularnewline
\midrule 
Trt & R1 & R2 & R3 & Overall & Exp & Ref & Diff (\% bias) & Coverage & Power\tabularnewline
\midrule
\midrule 
$A=1$ & 5 & 23 & 2 & 30 & \multirow{2}{*}{-2.04} & \multirow{2}{*}{-1.13} & \multirow{2}{*}{-0.91 (9\%)} & \multirow{2}{*}{0.92} & \multirow{2}{*}{0.79}\tabularnewline
\cmidrule{1-5} 
$A=0$ & 5 & 0 & 5 & 10 &  &  &  &  & \tabularnewline
\midrule 
$A=1$ & 5 & 17 & 3 & 25 & \multirow{2}{*}{-2.10} & \multirow{2}{*}{-1.31} & \multirow{2}{*}{-0.79 (21\%)} & \multirow{2}{*}{0.87} & \multirow{2}{*}{0.73}\tabularnewline
\cmidrule{1-5} 
$A=0$ & 5 & 0 & 10 & 15 &  &  &  &  & \tabularnewline
\midrule 
$A=1$ & 5 & 10 & 5 & 20 & \multirow{2}{*}{-2.14} & \multirow{2}{*}{-1.41} & \multirow{2}{*}{-0.73 (27\%)} & \multirow{2}{*}{0.83} & \multirow{2}{*}{0.70}\tabularnewline
\cmidrule{1-5} 
$A=0$ & 5 & 0 & 15 & 20 &  &  &  &  & \tabularnewline
\midrule 
$A=1$ & 5 & 3 & 7 & 15 & \multirow{2}{*}{-2.20} & \multirow{2}{*}{-1.54} & \multirow{2}{*}{-0.66 (34\%)} & \multirow{2}{*}{0.75} & \multirow{2}{*}{0.66}\tabularnewline
\cmidrule{1-5} 
$A=0$ & 5 & 0 & 20 & 25 &  &  &  &  & \tabularnewline
\bottomrule
\end{tabular}
\par\end{centering}
\caption{Multiple Imputation with a Mixture of Missing Data Types}
\end{table}

\begin{table}[H]
\begin{centering}
\begin{tabular}{cccccccccc}
\toprule 
 & \multicolumn{4}{c}{Missing Rate, \%} & \multicolumn{5}{c}{Trimmed Means + MI}\tabularnewline
\midrule 
Trt & R1 & R2 & R3 & Overall & Exp & Ref & Diff (\% bias) & Coverage & Power\tabularnewline
\midrule
\midrule 
$A=1$ & 5 & 23 & 2 & 30 & \multirow{2}{*}{-2.17} & \multirow{2}{*}{-1.14} & \multirow{2}{*}{-1.03 (-3\%)} & \multirow{2}{*}{0.92} & \multirow{2}{*}{0.88}\tabularnewline
\cmidrule{1-5} 
$A=0$ & 5 & 0 & 5 & 10 &  &  &  &  & \tabularnewline
\midrule 
$A=1$ & 5 & 17 & 3 & 25 & \multirow{2}{*}{-2.32} & \multirow{2}{*}{-1.31} & \multirow{2}{*}{-1.01 (-1\%)} & \multirow{2}{*}{0.93} & \multirow{2}{*}{0.88}\tabularnewline
\cmidrule{1-5} 
$A=0$ & 5 & 0 & 10 & 15 &  &  &  &  & \tabularnewline
\midrule 
$A=1$ & 5 & 10 & 5 & 20 & \multirow{2}{*}{-2.42} & \multirow{2}{*}{-1.42} & \multirow{2}{*}{-1.00 (-0\%)} & \multirow{2}{*}{0.94} & \multirow{2}{*}{0.88}\tabularnewline
\cmidrule{1-5} 
$A=0$ & 5 & 0 & 15 & 20 &  &  &  &  & \tabularnewline
\midrule 
$A=1$ & 5 & 3 & 7 & 15 & \multirow{2}{*}{-2.53} & \multirow{2}{*}{-1.54} & \multirow{2}{*}{-0.99 (1\%)} & \multirow{2}{*}{0.94} & \multirow{2}{*}{0.89}\tabularnewline
\cmidrule{1-5} 
$A=0$ & 5 & 0 & 20 & 25 &  &  &  &  & \tabularnewline
\bottomrule
\end{tabular}
\par\end{centering}
\caption{Trimmed Means + MI with a Mixture of Missing Data Types}
\end{table}

\subsection*{(e) Choice of $\alpha$ and Strong MNAR Assumption}

The analyst chooses $\alpha$- the proportion trimmed from each treatment
arm \textendash{} which can be set to any value above the maximum
proportion of missing data between the two arms. For all previous
simulations the value of $\alpha$ was chosen adaptively. Herein,
the adaptive choice of $\alpha$ is compared to a fixed choice where
$\alpha=0.5$. In addition, to investigate the strong MNAR assumption
of Theorem 1, the percent of missing values that would have fallen
below the trim point had they been observed is calculated for each
scenario. The adaptive and fixed $\alpha$ approaches are evaluated
under 10 different MNAR data generating mechanisms using the same
logit model as before. The $\alpha_{Y}$ parameter was set to values
of -0.5, -1, -1.5, -2, -2.5, -3, -4, -5, -7.5 and -10 while keeping
$\alpha_{0}=2.85$ and $\alpha_{A}=0$. Rates of missing data and
results are shown in Table 10.

\begin{table}[H]
\begin{centering}
\begin{tabular}{cccccccccc}
\toprule 
\multicolumn{2}{c}{Missing Rate, \%} & \multicolumn{4}{c}{Adaptive $\alpha$} & \multicolumn{4}{c}{Fixed $\alpha=0.5$}\tabularnewline
\midrule 
$A=1$ & $A=0$ & Diff (\% bias) & sMNAR & SE & MSE & Diff (\% bias) & sMNAR & SE & MSE\tabularnewline
\midrule
\midrule 
3 & 4 & -1.038 (-3.8\%) & 11.4\% & 0.321 & 0.104 & -1.015 (1.5\%) & 74.6\% & 0.355 & 0.126\tabularnewline
\midrule 
2 & 5 & -1.038 (-3.8\%) & 28.2\% & 0.313 & 0.100 & -1.012 (1.2\%) & 91.1\% & 0.352 & 0.124\tabularnewline
\midrule 
2 & 7 & -1.032 (3.2\%) & 45.9\% & 0.310 & 0.097 & -1.009 (0.9\%) & 97.0\% & 0.351 & 0.123\tabularnewline
\midrule 
3 & 8 & -1.025 (2.5\%) & 60.6\% & 0.309 & 0.096 & -1.008 (0.8\%) & 99.1\% & 0.351 & 0.123\tabularnewline
\midrule 
3 & 10 & -1.019 (1.9\%) & 70.3\% & 0.308 & 0.095 & -1.008 (0.8\%) & 99.6\% & 0.351 & 0.123\tabularnewline
\midrule 
3 & 11 & -1.015 (1.5\%) & 76.4\% & 0.308 & 0.095 & -1.007 (0.7\%) & 99.8\% & 0.350 & 0.123\tabularnewline
\midrule 
4 & 14 & -1.003 (0.3\%) & 83.8\% & 0.314 & 0.099 & -0.999 (0.1\%) & 100.0\% & 0.356 & 0.127\tabularnewline
\midrule 
5 & 15 & -1.002 (0.2\%) & 87.9\% & 0.312 & 0.098 & -1.003 (0.3\%) & 100.0\% & 0.355 & 0.126\tabularnewline
\midrule 
6 & 18 & -0.998 (0.2\%) & 93.4\% & 0.314 & 0.099 & -1.003 (0.3\%) & 100.0\% & 0.355 & 0.126\tabularnewline
\midrule 
7 & 20 & -0.995 (0.5\%) & 95.7\% & 0.313 & 0.098 & -1.001 (0.1\%) & 100.0\% & 0.351 & 0.123\tabularnewline
\bottomrule
\end{tabular}
\par\end{centering}
\caption{Comparison of Adaptive and Fixed $\alpha$}
\end{table}

Overall, both the fixed and adaptive $\alpha$ accurately estimate
the true difference in treatment effects $(\beta_{A}=-1)$. As $\alpha_{Y}$
moves further from 0, the percentage of missing values falling below
the trim point (i.e. sMNAR) increases. As a consequence, bias decreases
which is consistent with the theoretical result of Theorem 1. Unlike
in imputation, bias when using trimmed means is not directly related
to the fraction of missing data, but rather due to sMNAR. Thus, it
is possible to observe lower bias despite larger amounts of missing
data, as is demonstrated here. The fixed $\alpha$ approach consistently
has a higher sMNAR than the adaptive approach. This explains why using
the fixed \textgreek{a} has lower bias in all scenarios: the underlying
missing values have a higher likelihood of falling below the more
extreme trimming quantile. Using the adaptive $\alpha$ has a smaller
variance than the fixed $\alpha$ because the adaptive approach trims
the least amount of data possible. The smaller variance of the adaptive
approach translated to a smaller MSE than the fixed approach despite
the fixed approach having less bias. This simulation highlights the
bias vs variance tradeoff associated with increasing the percentage
of data trimmed.

\section{Application to a Clinical Trial}

We applied the methodologies described above to data from a double-blind
randomized clinical trial of two treatments (A and B) conducted in
patients with neuropathic pain due to diabetic neuropathy. Seventy-one
patients were randomized to treatment A and seventy to treatment B.
The outcome of interest was change in pain severity from baseline
to week 16, as assessed on a Visual Analog Scale (VAS). VAS is a well-studied
instrument for recording pain where a score of 100 reflects the \textquotedblleft worst
pain possible\textquotedblright{} and a score of 0 reflects \textquotedblleft no
pain\textquotedblright{}\textsuperscript{19}. Pain scores were recorded in a
digital diary daily by patients. At most, there were 16 weekly pain
measurements for each patient, produced by averaging daily pain recordings
during each week. 

\begin{table}[H]
\begin{centering}
\begin{tabular}{ccc}
\toprule 
Dropout Type & Treatment A & Treatment B\tabularnewline
\midrule
\midrule 
Adverse Events & 18 & 4\tabularnewline
Loss of Efficacy & 3 & 3\tabularnewline
Administrative & 12 & 14\tabularnewline
\midrule 
Total & 33 & 20\tabularnewline
\bottomrule
\end{tabular}
\par\end{centering}
\caption{Treatment Discontinuation Reasons}
\end{table}

Study discontinuation in this clinical trial was common, as there
were 53 (38\%) patients who did not stay on trial for 16 weeks. Discontinuation
differed among treatment arms, 33 (46\%) in the treatment A arm and
20 (29\%) in the treatment B arm. The reason for discontinuing the
study was recorded for each dropout and categorized as Adverse Event
(AE), Loss of Efficacy (LoE), or Administrative (Table 11). The rates
of study discontinuation, the time at which they occurred, and the
observed data before dropout were used to inform missing data assumptions.
AE and LoE generally occurred during the first half of the study period
while administrative dropout occurred uniformly throughout the trial.
On average AE and LoE occurred after 6.77 and 6.83 weeks on trial,
respectively, and administrative dropouts after 10.6 weeks. Based
on this exploratory data analysis and our clinical knowledge, we assume
that the dropouts classified as AE \& LoE were MNAR and administrative
dropouts were MCAR.

We applied 5 different methods to the trial data. First, the trimmed
means approach was applied globally to all dropouts (i.e. assumes
all dropouts MNAR). The fraction trimmed was chosen adaptively and
thus corresponded to the amount of dropout in the treatment A arm
(i.e. \textgreek{a} =0.46). To test the location shift assumption,
we performed a Kolmogorov-Smirnov test between the distribution of
trt A shifted by the treatment effect compared to the observed distribution
of treatment B. The test failed to reject that the untrimmed outcome
distributions were a location shift of one another (D=0.0946, p=0.9849).
We also applied multiple imputation to all dropouts in an ANOVA model
despite the MAR assumption being unlikely for many dropouts. Next,
we applied the approach that combines trimmed means and multiple imputation.
To do this, we trimmed AE and LoE (MNAR) and imputed administrative
dropout data (MCAR). As a consequence, the fraction trimmed was reduced
to \textgreek{a} = 0.30 in each of the imputed datasets. Lastly, we
applied two more approaches for historical reference, a complete case
analysis of all patients completing the trial (i.e. assumes all dropouts
MCAR) and a Last Observation Carried Forward (LOCF) analysis. 

\begin{table}[H]
\begin{centering}
\begin{tabular}{ccccc}
\toprule 
Method & Pain Difference & SE & 95\% CI & $p$-value\tabularnewline
\midrule
\midrule 
Trimmed Means & -14.48 & 7.61 & {[}-29.38, 0.43{]} & 0.055\tabularnewline
\midrule 
Trimmed Means + MI & -12.67 & 6.21 & {[}-24.83, -0.49{]} & 0.041\tabularnewline
\midrule 
Complete Case Analysis & -3.74 & 5.39 & {[}-14.45, 6.97{]} & 0.497\tabularnewline
\midrule 
Multiple Imputation & -2.71 & 4.68 & {[}-11.88, 6.45{]} & 0.537\tabularnewline
\midrule 
LOCF & -1.76 & 4.20 & {[}-10.06, 6.54{]}  & 0.675\tabularnewline
\bottomrule
\end{tabular}
\par\end{centering}
\caption{Clinical Trial Analysis Results}
\end{table}

Table 12 contains the results of each of these approaches. The trimmed
means applied to all dropouts showed the largest treatment difference
of -14.48 points lower on the pain VAS; however, the trimming inflated
the standard error to 7.61. Similarly, the combination of trimmed
means and MI had the second largest effect size -12.67. Contrarily,
the combination approach trimmed less data which resulted in a less
inflated standard error. These two methods, which involve trimming,
resulted in a larger effect size than the other methods because they
allow for the fact that the worse performing treatment had a higher
dropout rate. The other approaches presented are not appropriate given
our missing data assumptions, but were included for illustrative purposes
as a reference, especially the relative standard errors. 

Focusing on the SE column, we see that LOCF has the smallest standard
error, 4.20, as expected since the method does not admit to missing
data; it replaces all missing data with the last available data point,
in time, for each patient based on a single imputation. Some object
to this method being unrealistic in this situation. The next smallest
entry, 4.68, Multiple Imputation, treats all missing data as missing
at random, which may not be plausible for all discontinuation reasons
in this trial. Once again, the method creates data whenever they are
missing, except not once, as in LOCF, but multiple times in order
to better reflect the uncertainty induced by the missing data. The
complete case analysis has the next smallest value, 5.39. This was
only included for completeness. In general, the shortcomings of this
method are well known. The next smallest is the Trimmed Means + MI,
6.21. This approach achieves an unbiased comparison by maintaining
an \textquotedblleft equal percentage\textquotedblright{} of the data
for those who prematurely leave the study for cause. This method also
permits distinguishing between observations that are truly missing
at random and other observations for which the missing at random assumption
is not plausible (i.e. missing not at random). The last one, the Trimmed
Mean method has the largest standard error, 7.61. This is easily explained
because this is the method that discards the most data. This is excessive
in that the fraction of the missing data that are missing at random
are best handled as missing at random and thus amenable to multiple
imputation. 

\section{Discussion}

Our work extends the utility of the trimmed means approach for missing
data in two key ways: 1) It determines sufficient conditions for which
the trimmed means approach can identify the estimand for the population
treatment effect; and 2) It demonstrates that when different types
of missing data are present and can be distinguished, one could combine
the trimmed means approach with multiple imputation to improve estimation.

The trimmed means approach was originally designed to estimate a unique
estimand: the treatment difference in the best (100 \texttimes{} \textgreek{a})
\% of patients of each arm. The work herein allows us to view the
trimmed means approach in a different way, not a method estimating
a unique estimand, but a method that targets the usual estimand of
a clinical trial where accuracy depends on how well the assumptions
are satisfied. 

Missing data inferences are not possible without assumptions. The
strong MNAR assumption, similar to the MAR assumption, is untestable.
It is a conservative assumption that assumes every missing value falls
below the trimming quantile. It would be rare for this assumption
to hold perfectly; however, when applied to dropouts reporting loss
of efficacy, where poor outcomes are the primary cause of dropout,
the assumption may hold for enough of the missing data to justify
adopting the trimmed means approach. Also the numerical studies show
that the trimmed means estimator can still perform well under MNAR
scenarios that are not explicitly strong MNAR. Thus, the assumption
may be robust to slight deviations. One interesting paradox is that
while trimming more data leads to a loss in efficiency, theoretically
it allows the strong MNAR assumption to become more plausible since
missing values are then more likely to be trimmed. The simulation
comparing a fixed to adaptive choice of \textgreek{a} demonstrates
this. This bias/variance trade off should be considered when choosing
the value of \textgreek{a}. The location shift assumption may be more
realistic, is testable among the observed values, and is often assumed
in many statistical methods. In practice, the untrimmed fractions
of the distributions for each treatment arm of the study should be
compared using a Kolmogorov-Smirnov test to assess the validity of
this assumption. Of course should neither the location family nor
the strong MNAR assumptions be plausible, the trimmed means approach
can still be useful in estimating the original composite estimand
in the sub-population of the trial for which it was originally developed.

This research highlights the importance for administrators and physicians
conducting clinical trials to document the reasons for dropout. If
close collaboration between statisticians and clinicians can inform
which dropouts are MCAR, MAR, or MNAR then analysts may have a combination
of data that can be imputable and other data that should be trimmed
away using the trimmed means approach. This combination approach can
protect analysts from penalizing themselves using trimmed means globally
for all missing data but also respects the assumption that patients
may drop out of the trial due to poor health outcomes. If a fraction
of dropouts are missing because of factors unrelated to their unobserved
outcomes (MAR), the bias and loss of power using trimmed means for
all dropouts can be drastic. Choosing which missing data to treat
as MAR or MNAR will vary from trial to trial, and will also be dependent
on the particular estimand of interest. Clinical input is crucial
for the mixture approach to be effective. One limitation of the combination
approach is that MNAR dropouts must precede MAR dropouts; otherwise
the complete cases leveraged for imputations may have a different
outcome distribution than the MAR outcomes.

The clinical trial example highlights the importance of combining
trimmed means and multiple imputation. MCAR data is by definition
evenly distributed between each arm of the study; therefore, it does
not cause bias when applying trimmed means. However, there is considerable
power loss as shown in MCAR situation of our simulations (section
4a). In the clincal trial analysis, applying trimmed means to all
dropouts produced a larger estimate of the difference in treatment
effects, but because 46\% of observations were missing in one arm
of the study the standard errors were inflated, making it harder to
reject the null hypothesis of no difference between treatments. The
combination approach, however, preserved a similar estimate of the
treatment comparison and did not inflate standard errors as drastically.
The combination approach leverages a larger effective sample size
than applying trimmed means alone. 

The trimmed means approach is a creative solution to estimating treatment
effects in a clinical trial when missing data can safely be assumed
to be due to poor outcomes. As is the case in any missing data analysis,
especially those with MNAR data, no analytical method replaces a good
sensitivity analysis to determine the plausible range of what could
have happened. While no method can fully or confidently rectify the
issues caused by missing data, a combination of multiple imputation
and/or trimmed means could be useful when the assumptions of both
methods are satisfied. 

\begin{flushleft}
\textbf{Acknowledgements}
\par\end{flushleft}
We thank Jaffer Zaidi and Zack McCaw for useful discussions regarding the proofs of Theorems 1 and 2, respectively. Alex Ocampo was supported by NIH-5T32AI007358.

\begin{flushleft}
\textbf{References}
\par\end{flushleft}
\begin{enumerate}
\item Little R, Rubin DB. \textit{Statistical Analysis with
Missing Data}. 2nd. New York, NY: John Wiley \& Sons; 2002.

\item Rubin DB. Inference and missing data. \textit{Biometrika}. 1976;63(3):581-592. 

\item Rubin DB. \textit{Multiple Imputation for Nonresponse in Surveys}. Hoboken, NJ: John Wiley \& Sons; 2004. 

\item Robins JM, Rotnitzky A, Zhao LP. Analysis
of semiparametric regression models for repeated outcomes in the presence of missing data. \textit{Journal of the American Statistical Association}. 1995;90(429):106-121. 

\item Committee on National Statistics Division of Behavioral and Social Sciences and Education. National Research Council. \textit{The prevention and treatment of missing
data in clinical trials}. Washington, DC: National Academies Press; 2010.

\item Little, RJ. Comments on: Missing data methods
in longitudinal studies: A review. \textit{Test}. 2009;18(1):47-50. 

\item Wang M, Tchetgen-Tchetgen EJ. On varieties
of doubly robust estimators under missingness not at random with a shadow variable. \textit{Biometrika}. 2016;103(2):475-482. 

\item Carpenter, JR, Roger JH, Kenward MG. Analysis
of longitudinal trials with protocol deviation: a framework for relevant, accessible assumptions, and inference via multiple imputation. \textit{Journal of Biopharmaceutical Statistics}. 2013;23(6):1352-1371. 

\item Permutt T, Li F. Trimmed means for symptom
trials with dropouts. \textit{Pharmaceutical statistics}. 2017;16(1):20-28. 

\item Mallinckrodt CH, Clark WS, David SR. Accounting
for dropout bias using mixed-effects models. \textit{Journal of Biopharmaceutical Statistics}. 2001;11(1-2):9-21. 

\item Wang M, Liu J, Molenberghs G, Mallinckrodt CH. An evaluation of the trimmed mean approach in clinical trials with dropout. \textit{Pharmaceutical
Statistics}. 2018;17(3):278-289. 

\item Mehrotra DV, Liu F, Permutt T. Missing data in clinical trials: control-based mean imputation and sensitivity analysis. \textit{Pharmaceutical Statistics}. 2017;16(5):378-392. 

\item Good PI. Permutation, parametric, and bootstrap
tests of hypotheses. 3rd. New York, NY: Springer; 2004. 

\item Stigler SM. The asymptotic distribution of the trimmed mean. \textit{The Annals of Statistics}. 1973;1(3):472-477. 

\item Committee for Human Medicinal Products. ICH E9 (R1) addendum on estimands and sensitivity analysis in clinical trials to the guideline on statistical principles for clinical trials, Step 2b. London: European Medicines Evaluation Agency; Published August 30, 2017. Accessed June 15, 2018.

\item Cole SR, Frangakis CE. The
consistency statement in causal inference: a definition or an assumption?. \textit{Epidemiology}. 2009;20(1):3-5. 

\item Van Buuren, S, Groothuis-Oudshoorn K. mice:
Multivariate imputation by chained equations in R. \textit{Journal of Statistical Software}. 2010:1-68. 

\item Mohan K, Pearl J, Tian J. Graphical
models for inference with missing data. In: Mozer MC, ed. \textit{Advances in
Neural Information Processing Systems}. Cambridge, MA: The MIT Press; 2013. 

\item Burkhart B, Lorenz J. Pain measurement in man: Neurophysiological correlates
of pain. \textit{Electroencephalography and Clinical Neurophysiology}. 1984;107(4):227-253
\end{enumerate}

\pagebreak{}
\begin{center}
\textbf{Appendix: Proofs}
\par\end{center}
\begin{flushleft}
\textbf{Proof of Theorem 1.}
\end{flushleft}
Consider counterfactuals $Y_{1}$ and $Y_{0}$ with absolutely continuous
distribution functions $f_{1}(y)$ and $f_{0}(y)$ respectively both
defined over a common domain $(-\infty,\infty)$. Here, $R_{a}$ is
a binary indicator of $Y_{a}$ being observed or missing. Here, binary
treatment $a\in\{0,1\}$ determines which of the two counterfactuals
is observed and is intervened on through randomization. Lastly consider
the transformation of counterfactual $Y_{a}$ such that:

\[
U_{a}=\begin{cases}
Y_{a} & \text{if}\;R_{a}=1\\
min(\mathbf{Y_{a}}|R_{a}=1)-\epsilon & \text{if}\;R_{a}=0
\end{cases}\text{for}\;\epsilon>0
\]
There exist two sufficient conditions in order to prove the equality
of the trimmed means estimand and treatment difference in the whole
population. The first condition \textendash{} the location family
assumption - is that the distribution of potential outcomes from the
experimental group $Y_{1}\sim f_{1}(y)$ is in the same location family
as the distribution of potential outcomes from the reference group
$Y_{0}\sim f_{0}(y)$. Consider some constant $\Delta$ then$f_{0}(y)=f_{1}(y+\Delta)$.
If two distributions are a location shift of one another then $E[Y_{1}]=E[Y_{0}]+\Delta$
because the mean is the location parameter of a distribution. Also,
note that as a consequence all quantiles of these distributions are
a location shift of one another i.e. $F_{0}^{-1}(\alpha)+\Delta=F_{1}^{-1}(\alpha)$.
The second condition \textendash{} strong MNAR assumption - is that
all missing values fall below the point at which the distributions
are trimmed, i.e. $Y_{a}|R_{a}=0<F_{a}^{-1}(\alpha)$. The strong
MNAR assumption ensures that the composite outcome $U_{a}$ is trimmed
at the same value as $Y_{a}$ for all percentiles above the maximum
rate of missing data between the two arms, i.e. $F_{U_{a}}^{-1}(\alpha)=F_{a}^{-1}(\alpha)\forall\alpha:\alpha>Pr[R_{a}=0]$.
Leveraging both assumptions we can demonstrate the equality of the
two estimands:

\begin{align*}
\hspace{-7mm}E[U_{1}|U_{1}>F_{U_{1}}^{-1}(\alpha)]-E[U_{0}|U_{0}>F_{U_{0}}^{-1}(\alpha)] & =E[Y_{1}|Y_{1}>F_{1}^{-1}(\alpha)]-E[Y_{0}|Y_{0}>F_{0}^{-1}(\alpha)]\\
 & =\frac{1}{1-\alpha}\int_{F_{1}^{-1}(\alpha)}^{\infty}yf_{1}(y)dy-\frac{1}{1-\alpha}\int_{F_{0}^{-1}(\alpha)}^{\infty}yf_{0}(y)dy\\
 & =\frac{1}{1-\alpha}\left[\int_{F_{1}^{-1}(\alpha)}^{\infty}yf_{1}(y)dy-\int_{F_{0}^{-1}(\alpha)}^{\infty}yf_{0}(y)dy\right]\\
 & =\frac{1}{1-\alpha}\left[\int_{F_{1}^{-1}(\alpha)}^{\infty}yf_{1}(y)dy-\int_{F_{0}^{-1}(\alpha)}^{\infty}yf_{1}(y+\Delta)dy\right]\\
 & =\frac{1}{1-\alpha}\left[\int_{F_{1}^{-1}(\alpha)}^{\infty}yf_{1}(y)dy-\int_{F_{1}^{-1}(\alpha)}^{\infty}(x-\Delta)f_{1}(x)dx\right]\;x=y+\Delta\\
 & =\frac{1}{1-\alpha}\left[\int_{F_{1}^{-1}(\alpha)}^{\infty}yf_{1}(y)dy-\int_{F_{1}^{-1}(\alpha)}^{\infty}xf_{1}(x)dx+\Delta\int_{F_{1}^{-1}(\alpha)}^{\infty}f_{1}(x)dx\right]\\
 & =\frac{1}{1-\alpha}\left[\Delta\int_{F_{1}^{-1}(\alpha)}^{\infty}f_{1}(x)dx\right]\\
 & =\frac{1}{1-\alpha}\left[\Delta\left[1-\alpha\right]\right]\\
 & =\Delta\\
 & =E[Y_{0}]-E[Y_{0}]+\Delta\\
 & =E[Y_{1}]-E[Y_{0}]
\end{align*}

Which completes the proof.

$\;$

\textbf{Proof of Theorem 2.}

$\;$

Consider two distributions $F_{1}(\cdot)$ and $F_{0}(\cdot)$ which
are absolutely continuous distribution functions defined over a common
domain $(-\infty,\infty)$. Also, both distributions have an expectation
and that expectation is finite. Assume that the differences between
the $\alpha$-trimmed means are the same constant $\Delta\in\mathbb{R}$
for all $\alpha\in[0,1]$ that is:

\begin{align*}
E[Y_{1}|Y_{1}>F_{1}^{-1}(\alpha)]-E[Y_{0}|Y_{0}>F_{0}^{-1}(\alpha)] & =\Delta\\
E[Y_{0}|Y_{0}>F_{0}^{-1}(\alpha)] & =E[Y_{1}|Y_{1}>F_{1}^{-1}(\alpha)]-\Delta\\
\frac{1}{1-\alpha}\int_{F_{0}^{-1}(\alpha)}^{\infty}yf_{0}(y)dy & =\frac{1}{1-\alpha}\int_{F_{1}^{-1}(\alpha)}^{\infty}yf_{1}(y)dy-\Delta\\
\frac{1}{1-\alpha}\int_{F_{0}^{-1}(\alpha)}^{\infty}yf_{0}(y)dy & =\frac{1}{1-\alpha}\int_{F_{1}^{-1}(\alpha)}^{\infty}yf_{1}(y)dy-\frac{(1-\alpha)}{(1-\alpha)}\Delta\\
\frac{1}{1-\alpha}\int_{F_{0}^{-1}(\alpha)}^{\infty}yf_{0}(y)dy & =\frac{1}{1-\alpha}\int_{F_{1}^{-1}(\alpha)}^{\infty}yf_{1}(y)dy-\frac{1}{(1-\alpha)}\Delta\int_{F_{1}^{-1}(\alpha)}^{\infty}f_{1}(y)dy\\
\frac{1}{1-\alpha}\int_{F_{0}^{-1}(\alpha)}^{\infty}yf_{0}(y)dy & =\frac{1}{1-\alpha}\left[\int_{F_{1}^{-1}(\alpha)}^{\infty}yf_{1}(y)dy-\int_{F_{1}^{-1}(\alpha)}^{\infty}\Delta f_{1}(y)dy\right]\\
\frac{1}{1-\alpha}\int_{F_{0}^{-1}(\alpha)}^{\infty}yf_{0}(y)dy & =\frac{1}{1-\alpha}\left[\int_{F_{1}^{-1}(\alpha)}^{\infty}(y-\Delta)f_{1}(y)dy\right]\\
\frac{1}{1-\alpha}\int_{F_{0}^{-1}(\alpha)}^{\infty}yf_{0}(y)dy & =\frac{1}{1-\alpha}\left[\int_{F_{1}^{-1}(\alpha)-\Delta}^{\infty}xf_{1}(x+\Delta)dx\right]\;\;x=y-\Delta,\;dx=dy\\
\int_{F_{0}^{-1}(\alpha)}^{\infty}yf_{0}(y)dy & =\int_{F_{1}^{-1}(\alpha)-\Delta}^{\infty}xf_{1}(x+\Delta)dx
\end{align*}

At this stage perform the substitution of $y=F_{0}^{-1}(\beta),\;d\beta=f_{0}(y)dy$
for the integral on the left side of the equation and $x=F_{1}^{-1}(\beta)-\Delta,\;d\beta=f(x+\Delta)dx$
for the integral on the right side of the equation such that:

\begin{align*}
\int_{\alpha}^{1}F_{0}^{-1}(\beta)d\beta & =\int_{\alpha}^{1}[F_{1}^{-1}(\beta)-\Delta]d\beta\\
\frac{d}{d\alpha}\int_{\alpha}^{1}F_{0}^{-1}(\beta)d\beta & =\frac{d}{d\alpha}\int_{\alpha}^{1}[F_{1}^{-1}(\beta)-\Delta]d\beta\\
-F_{0}^{-1}(\alpha) & =-[F_{1}^{-1}(\alpha)-\Delta],\;\;\alpha\in(0,1)\\
F_{0}^{-1}(\alpha) & =F_{1}^{-1}(\alpha)-\Delta
\end{align*}

From here it follows that $F_{1}-\Delta=F_{0}$ and it is proven that
$F_{1}$ is a location shift of $F_{0}$ almost everywhere on the
domain i.e. $f_{0}(y)=f_{1}(y+\Delta)$.
\end{document}